\documentclass[aps,reprint,superscriptaddress]{revtex4-1}
\usepackage{amsmath,bm}
\usepackage{amssymb}
\usepackage{color}
\usepackage{braket}
\usepackage{multirow}
\usepackage[colorlinks,bookmarks=false,citecolor=red,linkcolor=blue,urlcolor=blue]{hyperref}
\usepackage{graphicx}
\usepackage{mathrsfs,bbm}
\usepackage[caption=false]{subfig}
\usepackage{wrapfig}
\usepackage[version=3]{mhchem}
\usepackage{amsfonts}
\usepackage{etoolbox}

\def \dag {^{\dagger}}
\renewcommand{\i}{{\mathrm{i}}}
\def \dd  {{\partial}}

\begin{document}

\title{Level crossings induced by a longitudinal coupling in the transverse field Ising chain}
\author{Gr\'egoire Vionnet}
\affiliation{Institute of Physics, \'Ecole Polytechnique F\'ed\'erale de Lausanne (EPFL), CH-1015 Lausanne, Switzerland}
\author{Brijesh Kumar}
\affiliation{School of Physical Sciences, Jawaharlal Nehru University (JNU), New Delhi 110067, India.}
\author{Fr\'ed\'eric Mila}
\affiliation{Institute of Physics, \'Ecole Polytechnique F\'ed\'erale de Lausanne (EPFL), CH-1015 Lausanne, Switzerland}

\date{\today}
\begin{abstract}
We study the effect of 
antiferromagnetic longitudinal coupling on the one-dimensional transverse 
field Ising model with nearest-neighbour couplings. In the topological phase 
where, in the thermodynamic limit, the ground state is twofold degenerate, 
we show that, for a finite system of $N$ sites, the longitudinal coupling 
induces $N$ level crossings between the two lowest lying states as a function 
of the field. We also provide strong arguments suggesting that these $N$ 
level crossings all appear simultaneously as soon as the longitudinal coupling 
is switched on. This conclusion is based on perturbation theory, and a mapping 
of the problem onto the open Kitaev chain, for which we write down the 
complete solution in terms of Majorana fermions.
\end{abstract}

\maketitle

The topological properties of matter are currently attracting a considerable 
attention~\cite{kane,zhang}. One of the hallmarks of a topologically non 
trivial phase is the presence of surface states. In one dimension, the first 
example was the spin-1 chain that was shown a long time ago to have a
gapped phase~\cite{haldane} with two quasi-degenerate low-lying states 
(a singlet and a triplet) on open chains~\cite{kennedy}. These low-lying 
states are due to the emergent spin-1/2 degrees of freedom at the edges of 
the chains which combine to make a singlet ground state with an almost 
degenerate low-lying triplet for an even number of sites, 
and a triplet ground state with an almost degenerate low-lying singlet when 
the number of sites is odd. In that system, the emergent degrees of freedom 
are magnetic since they carry a spin 1/2, and they can be detected by standard 
probes sensitive to local magnetisation such as NMR~\cite{tedoldi}. 

In fermionic systems, a topological phase is present if the model includes a 
pairing term (as in the mean-field treatment of a p-wave superconductor), 
and the emergent degrees of freedom are two Majorana fermions localised at 
the opposite edges of the chain~\cite{kitaev}. Their detection is much less 
easy than that of magnetic edge states, and it relies on indirect
consequences such as their impact on the local tunneling density of 
states~\cite{mourik,nadj-perge}, or the presence of two quasi-degenerate 
low-lying states in open systems. In that respect, it has been suggested to 
look for situations where the low-lying states cross as a function of an 
external parameter, for instance the chemical potential, to prove that there 
are indeed two low-lying states~\cite{sarma}. 

In a recent experiment with chains of Cobalt atoms evaporated onto a 
Cu${}_2$N/Cu(100) substrate~\cite{toskovic}, the presence of level crossings 
as a function of the external magnetic field has been revealed by scanning 
tunneling microscopy, which exhibits a specific signature whenever the 
ground state is degenerate. The relevant effective model for that system is 
a spin-1/2 XY model in an in-plane magnetic field. The exact diagonalisation 
of finite XY chains has indeed revealed the presence of 
quasi-degeneracy between the two lowest energy states, that are well separated 
from the rest of the spectrum, and a series of level crossings between them
as a function of the magnetic field~\cite{dmitriev}.  Furthermore, the 
position of these level crossings is in good agreement with the experimental 
data. It has been proposed that
these level crossings are analogous to those predicted in topological 
fermionic spin chains, and that they can be interpreted as a consequence 
of the Majorana edge modes~\cite{mila}. 

The topological phase of the XY model in an in-plane magnetic field is 
adiabatically connected to that of the transverse field Ising model, in 
which the longitudinal spin-spin coupling (along the field) is switched off. 
However, in the transverse field Ising model, the two low-lying states never 
cross as a function of the field, as can be seen from the magnetisation curve 
calculated by Pfeuty a long time ago~\cite{pfeuty}, and which does not show 
any anomaly. The very different behaviour of the XY model in an in-plane field 
in that respect calls for an explanation. The goal of the present paper is to 
provide such an explanation, and to show that the presence of $N$ level 
crossings, on a chain of $N$ sites, is generic as soon as an antiferromagnetic 
longitudinal coupling is switched on. To achieve this goal, we have studied 
a Hamiltonian which interpolates between the exactly solvable transverse field 
Ising (TFI) and the longitudinal field Ising (LFI) chains.  
The approach that best accounts for these level crossings turns out to  be an 
approximate mapping onto the exactly solvable Kitaev chain, which contains all 
the relevant physics. In the Majorana representation, the level crossings are 
due to the interaction between Majorana fermions localised at each end of the 
chain.

The paper is organized as follows. In section \ref{sec:model}, we present the 
model and give some exact diagonalisation results on small chains to get an 
intuition of the qualitative behaviour of the spectrum. We show in section 
\ref{sec:PT} that  perturbation theory works in principle but is rather 
limited because of the difficulty to go to high order. We then turn to an 
approximate mapping onto the open Kitaev chain via a mean-field decoupling in 
section \ref{sec:MF}. The main result of this paper is presented in section 
\ref{sec:Maj}, namely the explanation of the level crossings in a Majorana 
representation. Finally, we conclude with a a quick discussion of some 
possible experimental realisations in section \ref{sec:exp}.

\section{Model} \label{sec:model}
We consider the transverse field spin-1/2 Ising model with an additional 
antiferromagnetic longitudinal spin-spin coupling along the field, i.e. the 
Hamiltonian
\begin{equation} \label{eqHspin}
H=J_x \sum_{i=1}^{N-1}S_i^x S^x_{i+1} + J_z \sum_{i=1}^{N-1}S^z_{i} S^z_{i+1} 
- h \sum_{i=1}^N S^z_{i}
\end{equation}
with $J_z \geq 0$~\footnote{This Hamiltonian is equivalent to an XY model in 
an in-plane magnetic field, but we chose to rotate the spins around the 
$x$-axis so that we recover the usual formulations of the TFI and LFI 
models as special cases.}. 
This model can be seen as an interpolation between the TFI model ($J_z=0$) 
and the LFI model ($J_x=0$). The case $J_z=J_x$ corresponds to the effective 
model describing the experiment in Ref.~\onlinecite{toskovic}, up to small 
irrelevant terms~\footnote{In Ref.~\onlinecite{toskovic} it is explained that 
the $\pm 3/2$ doublet of the spin 3/2 Cobalt adatoms can be 
projected out by a Schrieffer-Wolff transformation due to the strong 
magnetic anisotropy. The resulting effective spin 1/2 model is the one of 
equation \eqref{eqHspin} with $J_x=J_z$ and additional nearest-neighbour 
out-of-plane and next-nearest-neighbour in-plane Ising couplings. 
These additional terms do not lead to qualitative changes because the model is still
symmetric under a $\pi$-rotation of the spins around the 
$z$-axis, and since their coupling constants are small ($\sim 0.1J_x$) 
they have only a small quantitative effect in exact diagonalisation results.}. 
Since we will be mostly interested in the parameter range 
$0\leq J_z\leq J_x$, we will measure energies in units of $J_x$ by setting 
$J_x=1$ henceforth. The spectrum of the Hamiltonian in Eq.~\eqref{eqHspin} is 
invariant under $h\to -h$ since the Hamiltonian is invariant if we 
simultaneously rotate the spins around the $x$-axis so that 
$S^z_i \to -S^z_i~\forall i$. Hence, we will in most cases quote the results 
only for $h \geq 0$.

The TFI limit of $H$ can be solved exactly by Jordan-Wigner mapping onto a 
chain of spinless fermions~\cite{pfeuty}. In the thermodynamic limit, it is 
gapped with a twofold degenerate ground state for $h < h_c=1/2$, and undergoes
a quantum phase transition at $h=h_c$ to a non-degenerate gapped ground state 
for $h>h_c$. The twofold degeneracy when $h<h_c$ can be described by two 
zero-energy Majorana edge modes~\cite{kitaev}. As a small positive $J_z$ is 
turned on, there is no qualitative change in the thermodynamic limit, except 
that $h_c$ increases with $J_z$. Indeed, the model is then equivalent to the 
ANNNI model in a transverse field which has been extensively studied before, 
see for example \cite{chakrabarti,jalal}. A second order perturbation 
calculation in $1/h$ yields $h_c =1/2+(3/4)J_z+O(J_z^2) $ for small $J_z$ and 
$h_c=1/2+J_z +O(1/J_z)$ for large $J_z$~\cite{rujan,hassler}. Since, for 
$J_z \gtrsim 1$, there are other phases arising~\cite{hassler}, we shall 
mostly consider $J_z \lesssim 1$ in the following in order to stay in the 
phase with a degenerate ground state.

\begin{figure}[t]
{
	\includegraphics{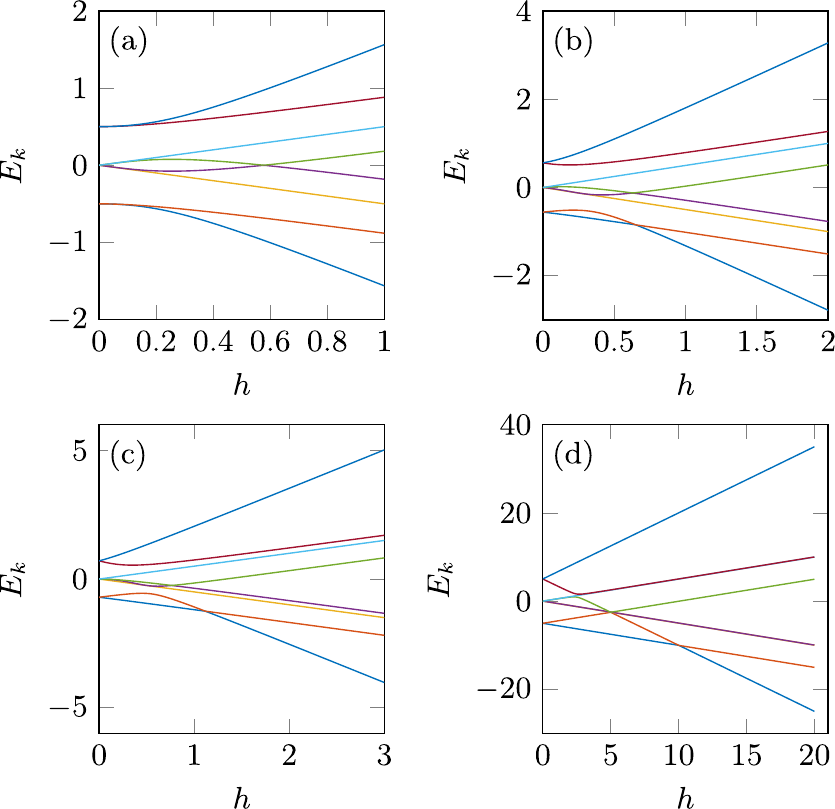}
}
\caption{Exact diagonalisation spectrum as a function of $h$ for $N=3$ with 
$J_z=0$~(a), $J_z=0.5$~(b), $J_z=1$~(c) and $J_z=10$~(d).
}
\label{fig1}
\end{figure}

For a finite size chain, the twofold degeneracy of the TFI model at $0<h<h_c$ 
is lifted and there is a small non-vanishing energy splitting 
$\epsilon = E_1-E_0$ between the two lowest energy states,  where the 
$E_k$ are the eigenenergies and $E_k \leq E_{k+1} ~ \forall k$. This splitting 
is exponentially suppressed with the system length, 
$\epsilon \sim \exp(-N/\xi)$~\cite{kitaev}. These two quasi-degenerate states 
form a low energy sector separated from the higher energy states. The spectrum 
for $J_z=0$ and $N=3$ is shown in Fig.~\ref{fig1}a. For $J_z > 0$, the 
splitting $\epsilon$ has an oscillatory behaviour and vanishes for some values 
of $h$. For $N=3$, it vanishes once for $h>0$. See the spectrum for $J_z=0.5$ 
and $J_z=1$ in Figs~\ref{fig1}b-c. As $J_z$ becomes large, there is no low 
energy sector separated from higher energy states any more. In the LFI limit, 
$J_z\to\infty$, the eigenstates have a well defined magnetisation in the 
$z$-direction and the energies are linear as a function of $h$, see 
Fig.~\ref{fig1}d. In this limit, the level crossings are obvious.
As the field is increased, the more polarised states become favoured, which 
leads to level crossings.
\begin{figure}[ht]
{
	\includegraphics{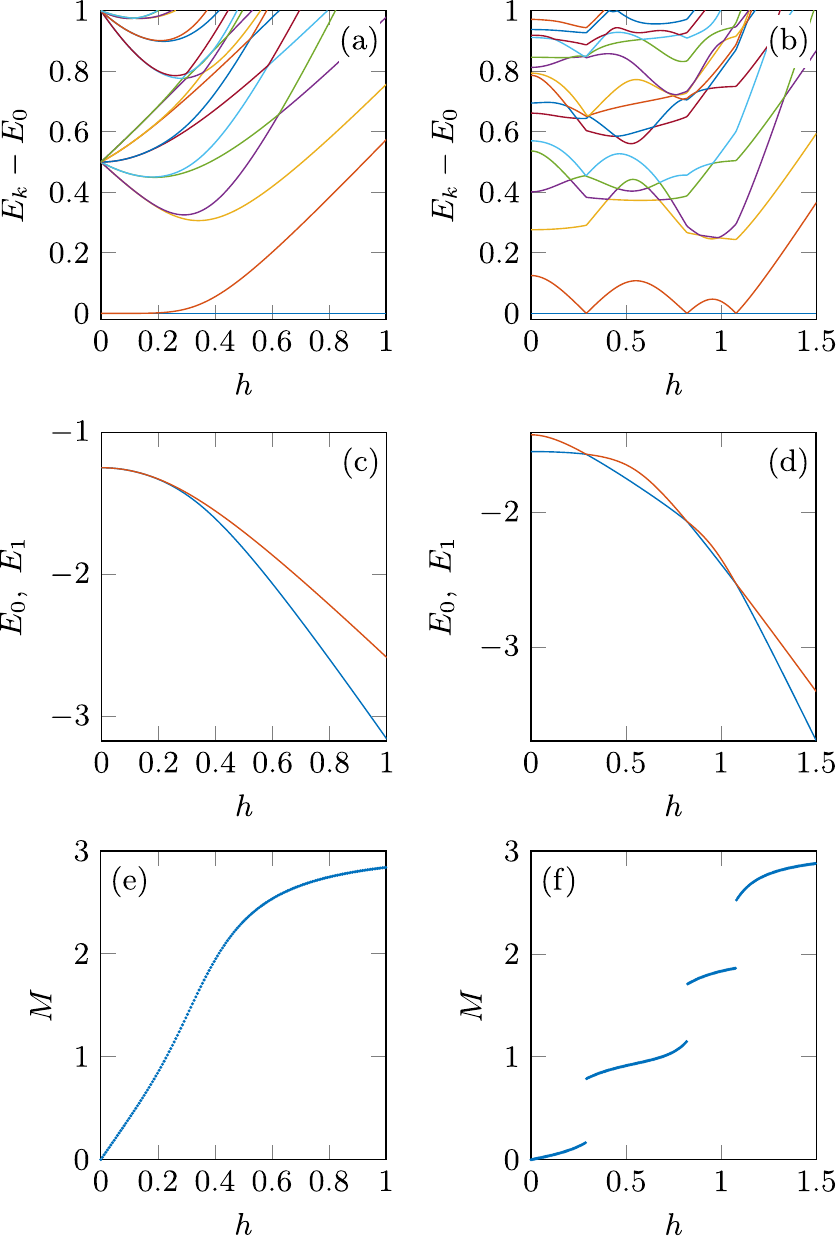}
}
\caption{Exact diagonalisation spectrum relative to the ground state energy, 
$E_k-E_0$, (a), two lowest energies $E_0$ and $E_1$ (c) and magnetisation 
$M=-\dd E_0/\dd h$ (e) as a function of $h$ for $N=6$ with 
$J_z=0$ (TFI limit). The plots in (b), (d) and (f) show the same for $J_z=0.75$.
}
\label{fig2}
\end{figure}

\begin{figure}
{
	\includegraphics{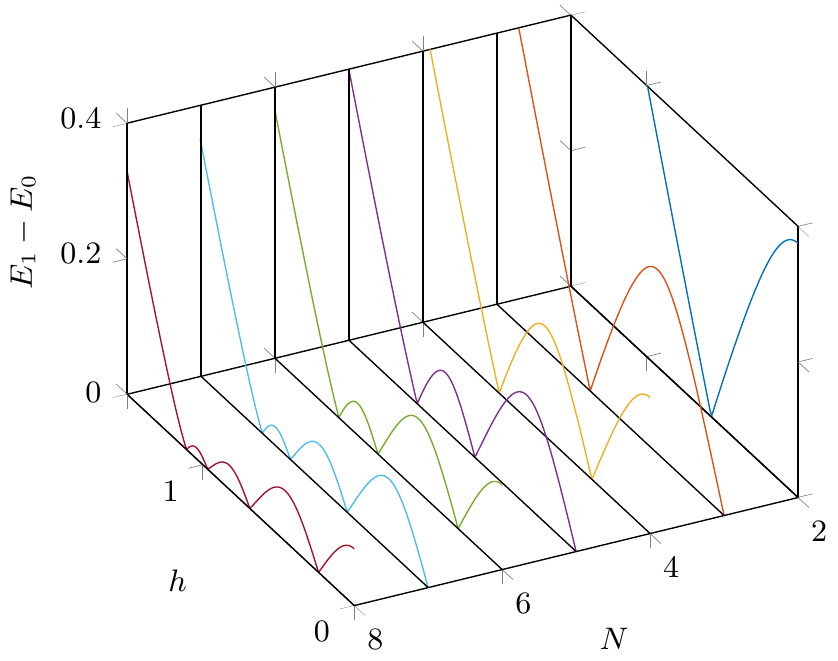}
}
\caption{Exact diagonalisation energy splitting $\epsilon=E_1-E_0$ between the 
two lowest energy states as a function of $h$ for several $N$ and $J_z=0.75$. 
There are $N$ level crossings since the spectrum is symmetric under $h\to -h$.
}
\label{fig3}
\end{figure}

The plots in Fig.~\ref{fig1} are instructive for very small $N$ but become 
messy for larger chains. In Figs \ref{fig2}a-b, we show the spectrum relative 
to the ground state energy, i.e. $E_k-E_0$, of a chain of $N=6$ sites for 
$J_z=0$ and $J_z=0.75$. The energies $E_0$ and $E_1$ are plotted in 
Figs~\ref{fig2}c-d for the same parameters. The structure of the spectrum is 
similar to the $N=3$ case, except that now $\epsilon$ vanishes at three points 
for $h>0$. In general, there are $N$ points of exact degeneracy where the 
splitting $\epsilon$ vanishes since the spectrum is symmetric under $h\to -h$. 
This is shown in Fig.~\ref{fig3} for $2\leq N\leq 8$. For $N$ even, there are 
$N/2$ level crossings for $h>0$, and for $N$ odd, there are $(N-1)/2$ level 
crossings for $h>0$ and one at $h=0$. 

\begin{figure}[t]
{
	\includegraphics{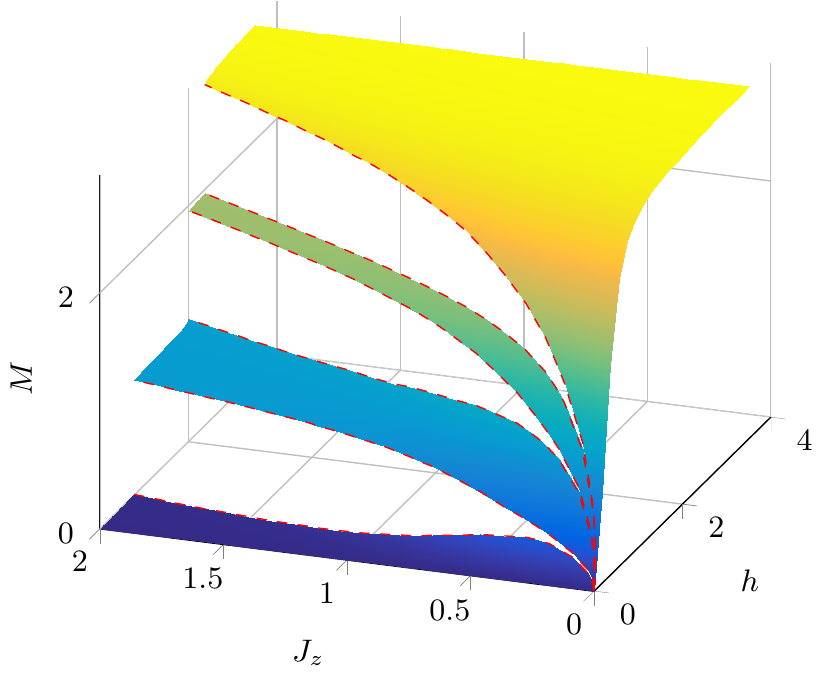}
}
\caption{Exact diagonalisation magnetisation $M$ as a function 
of $h$ and $J_z$ for $N=6$. The red dashed lines indicate the discontinuities.
}
\label{fig4}
\end{figure}

As shown in Figs \ref{fig2}e-f, the level crossings lead to jumps in the 
magnetisation $M(h)=-\dd E_0/\dd h$. The number of magnetisation 
jumps turns out to be independent of $J_z$ for $0<J_z<\infty$, as illustrated 
in Fig.~\ref{fig4}. In the LFI limit, most of the jumps merge together at 
$h=J_z$, with an additional jump persisting for even $N$ at 
$h=J_z/2$~\footnote{In the LFI model the lowest energy with a given 
magnetisation \mbox{$M=\sum_iS_i^z$} is $E_{0,M= 0} =-J_z(N-1)/4$ and 
$E_{0,M\neq 0} = E_{0,0}+ J_z(|M|-1/2) - Mh$. Thus for even $N$, the ground 
state has $M=0$ for $0<h<J_z/2$, $M=1$ for $J_z/2<h<J_z$ and $M=N/2$ for 
$h>J_z$, whereas for odd $N$ the ground state has $M=1/2$ for $0<h<J_z$ 
and $M=N/2$ for $h>J_z$.}. In this large $J_z$ region, however, there is no 
quasi-degeneracy and the magnetisation jumps indicate level crossings but no 
oscillation in contrast to the small $J_z$ region. Since there are no level 
crossings in the TFI limit, one might expect the number of crossings to 
decrease as $J_z$ decreases. However, the exact diagonalisation results do 
not support this scenario, and hint to all level crossings appearing at the 
same time as soon as $J_z\neq 0$. This is a remarkable feature that we shall 
explain in the following.

A useful equivalent representation of the Hamiltonian in Eq.~\eqref{eqHspin} 
in terms of spinless fermions is obtained by applying the Jordan-Wigner 
transformation used to solve exactly the TFI model~\cite{pfeuty},
 \begin{equation} \label{eqJW}
	\begin{cases}
S_i^x = \frac{1}{2}(c_i\dag  + c_i)\exp\left(\i\pi\sum_{j<i}c_j\dag c_j\right)\\
S_i^y = \frac{1}{2\i}(c_i\dag - c_i)\exp\left(\i\pi\sum_{j<i}c_j\dag c_j\right)\\
S_i^z=c_i\dag c_i - \frac{1}{2},
	\end{cases}
\end{equation}
which yields
      \begin{equation}
	\begin{split} \label{eqHfermions}
		H &= \frac{1}{4}  \sum_{i=1}^{N-1}(c_i\dag-c_i)(c_{i+1}\dag+c_{i+1}) 
		- h \sum_{i=1}^N (c_i\dag c_i - \frac{1}{2})\\&\quad 
		+ J_z \sum_{i=1}^{N-1}(c_i\dag c_i 
		- \frac{1}{2})(c_{i+1}\dag c_{i+1} - \frac{1}{2})
	\end{split}
\end{equation}
where the $c_i,c_i\dag$ are fermionic annihilation and creation operators. 
This is the Hamiltonian of a spinless p-wave superconductor with 
nearest-neighbour density-density interaction. As for the simpler TFI model, 
the Hamiltonian is symmetric under a $\pi$-rotation of the spins around the 
$z$-axis, $S^x_{i}\to -S^x_{i}$ and $S^y_{i}\to -S^y_{i}$ in the spin 
language. This leads to two parity sectors given by the parity operator
\begin{equation}
	P=e^{\i \pi \sum_{j=1}^N c_j\dag c_j} = (-2)^N S_1^z\cdots S_N^z.
\end{equation}
In other words, the Hamiltonian does not mix states with even and odd number 
of up spins, or equivalently with even and odd number of fermions. The ground 
state parity changes at each point of exact degeneracy, and thus alternates as 
a function of the magnetic field for $J_z > 0$. This can be understood 
qualitatively by looking at Fig.~\ref{fig2}f. The magnetisation plateaus are 
roughly at $M=0,1,2,3$. Hence to jump from one plateau to the next, one spin 
has to flip, thus changing the sign of the parity $P$.


\section{Perturbation theory} \label{sec:PT}

As a first attempt to understand if the $N$ level crossings develop 
immediately upon switching on $J_z$, we treat the  
$V=J_z \sum_{i=1}^{N-1}S^z_{i} S^z_{i+1}$ term as a perturbation to the 
exactly solvable transverse field Ising model. One may naively expect that 
degenerate perturbation theory is required since the TFI chain has a 
quasi-twofold degeneracy at low field. Fortunately, the two low-energy states 
live in different parity sectors~\cite{pfeuty} that are not mixed by the 
perturbation $V$. We can therefore apply the simple Rayleigh-Schr\"odinger 
perturbation theory in the range of parameters we are interested in, i.e. 
$J_z \lesssim 1$.
\begin{figure}
{
	\includegraphics{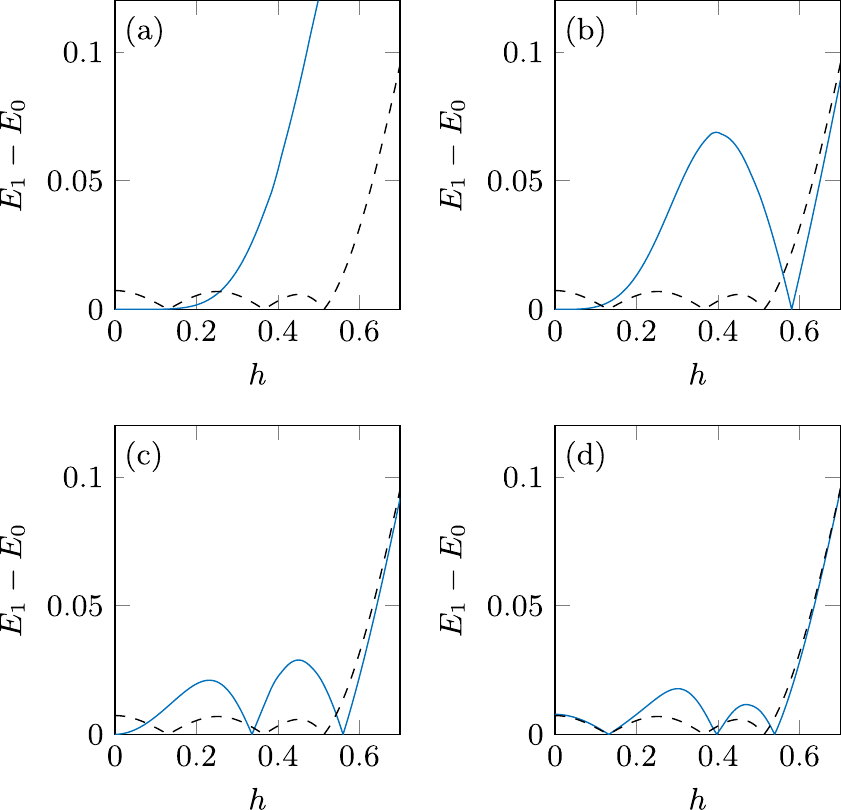}
}
\caption{Energy splitting at zeroth (a), first (b), second (c) and third (d) 
order (blue solid lines) compared to the exact diagonalisation result (black 
dashed lines) for $N=6$ and $J_z=0.25$. 
}
\label{fig5}
\end{figure}
\begin{figure}[t]
{
	\includegraphics{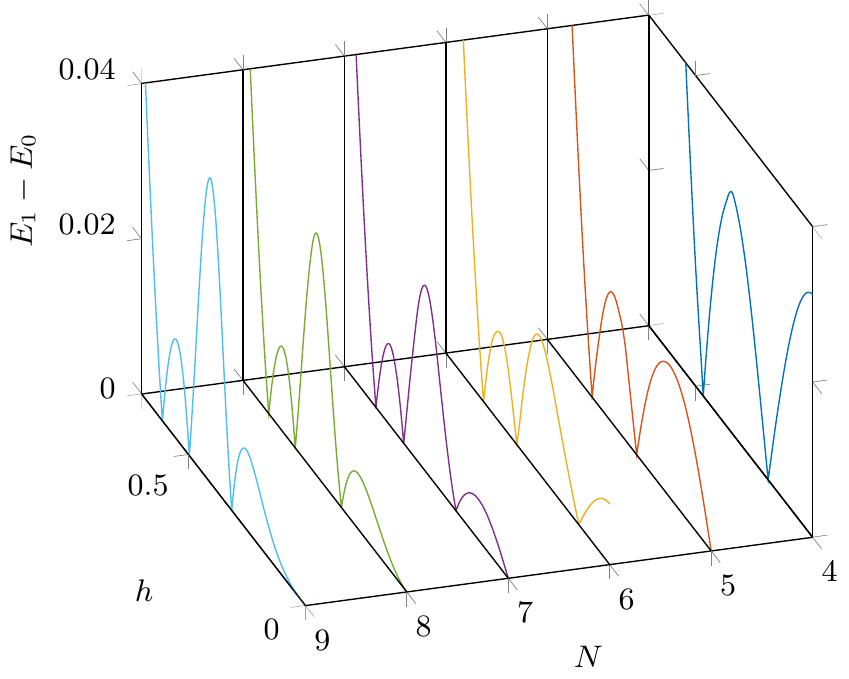}
}
\caption{Energy splitting at third order in perturbation theory for several 
$N$ and $J_z=0.25$. Since the spectrum is symmetric under $h\to -h$, there 
are $N$ level crossings for $N\leq 7$ and $7$ level crossings for $N \geq 7$.
}
\label{fig6}
\end{figure}

Writing $A_i=c_i\dag + c_i$ and $B_i=c_i\dag - c_i$, the perturbation can be 
rewritten as $V=({J_z}/{4}) \sum_{i=1}^{N-1}B_iA_iB_{i+1}A_{i+1}$.
The unperturbed eigenstates are $\ket{m} = \Upsilon\dag_m\ket{0}$ where 
$\ket{0}$ is the ground state and the $\Upsilon\dag_m$ are a product of the 
creation operators corresponding to the Bogoliubov fermions. 
The matrix elements are then
\begin{equation}
	\begin{split}
		\braket{n|V|m}=\frac{J_z}{4}\sum_{i=1}^{N-1}
		\braket{0|\Upsilon_n B_iA_iB_{i+1}A_{i+1}\Upsilon\dag_m|0}
	\end{split}
\end{equation}
which can be computed by applying Wick's theorem, similarly to how correlation 
functions are found in \cite{lieb}. We computed the effect of $V$ up to third 
order, with the basis of virtual states slightly truncated, namely by keeping 
states with at most three Bogoliubov fermions. Since the more fermions there 
are in a state, the larger its energy, we expect this approximation to be 
excellent.

As shown in Fig.~\ref{fig5}, the number of crossings increases with the order 
of perturbation, and to third order in perturbation, the results for $N=6$ 
sites are in qualitative agreement with exact diagonalisations. From the way 
level crossings appear upon increasing the order of perturbation theory, one 
can expect to induce up to $2m+1$ level crossings if perturbation theory
is pushed to order $m$, see Fig.~\ref{fig6}. So these results suggest that the 
appearance of level crossings is a perturbative effect, and that, for a given 
size $N$, pushing perturbation theory to high enough order will indeed lead to 
$N$ level crossings for small $J_z$. However, in practice, it is impossible to 
push perturbation theory to very high order. Indeed, the results at order 3 
are already very demanding. So, these pertubative results are encouraging, but 
they call for an alternative approach to actually prove that the number of 
level crossings is indeed equal to $N$, and that these level crossings appear 
as soon as $J_z$ is switched on.

\section{Fermionic mean-field approximation} \label{sec:MF}
\begin{figure}
{
	\includegraphics{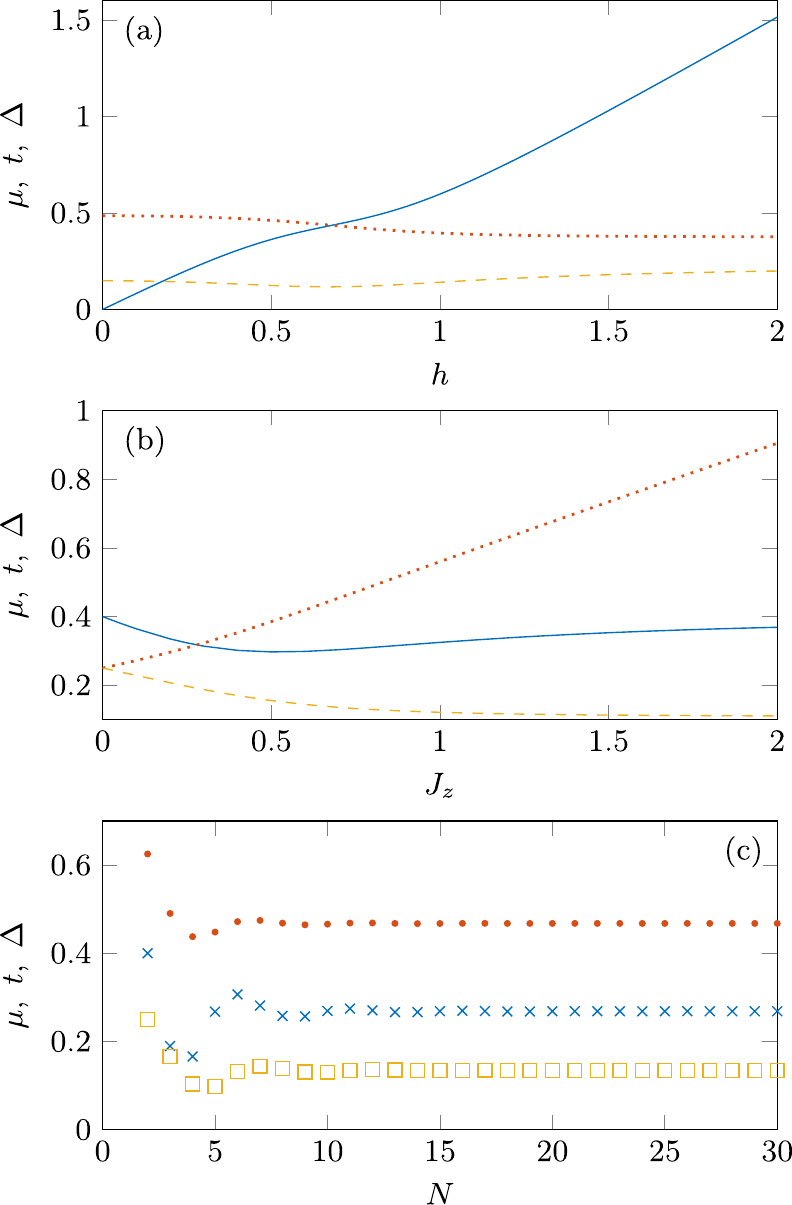}
}
\caption{Self-consistent mean-field parameters $\mu$ (blue solid lines and 
crosses), $t$ (red dotted lines and dots) and $\Delta$ (yellow dashed lines 
and squares) as a function of $h$ for $J_z=0.75$ and $N=6$ (a), as a function 
of $J_z$ for $h=0.4$ and $N=6$ (b) and as a function of $N$ for $h=0.4$ and 
$J_z=0.75$ (c).
}
\label{fig7}
\end{figure}

In the fermionic representation, Eq.~$\eqref{eqHfermions}$, there is a quartic 
term that cannot be treated exactly. Here, we approximate it by mean-field 
decoupling. In such an approximation, one assumes the system can be well 
approximated by a non-interacting system (quadratic in fermions) with 
self-consistently determined parameters. For generality, we decouple the 
quartic term in all three mean-field channels consistent with Wick's theorem,
\begin{equation}
	\begin{split}
		\label{eqMF}
 c\dag_{i}&c_{i}c\dag_{i+1}c_{i+1} \approx \\
 &\braket{c\dag_{i}c_{i}}	c\dag_{i+1}c_{i+1} + \braket{c\dag_{i+1}c_{i+1}} 	
 c\dag_{i}c_{i} - \braket{c\dag_{i}c_{i}} \braket{c\dag_{i+1}c_{i+1}} 	\\
- &\braket{c\dag_{i}c\dag_{i+1}} c_{i}c_{i+1} - \braket{c_{i}c_{i+1}}
c\dag_{i}c\dag_{i+1}  +\braket{c\dag_{i}c\dag_{i+1}} \braket{c_{i}c_{i+1}} \\ 
+&\braket{c\dag_{i}c_{i+1}} c_{i}c\dag_{i+1} +\braket{c_{i}c\dag_{i+1}} 
c\dag_{i}c_{i+1}  - \braket{c\dag_{i}c_{i+1}} \braket{c_{i}c\dag_{i+1}}.
	\end{split}
\end{equation}
Here, $\braket{.}$ denotes the ground state expectation value.
The $3N-2$ self-consistent parameters $\braket{c\dag_{i}c_{i}}$, 
$\braket{c\dag_{i}c\dag_{i+1}}$ and $\braket{c\dag_{i}c_{i+1}}$ can be found 
straightforwardly by iteratively solving the quadratic mean-field Hamiltonian.

As it turns out, it is more instructive to consider only three self-consistent 
parameters. To do so, we solve the mean-field approximation of the 
translationally invariant Hamiltonian ($c_{N+1}=c_1$), 
\begin{equation}\begin{split}
	H' &= \sum_{i=1}^{N} \left\lbrace \frac{1}{4}  
	(c_i\dag-c_i)(c_{i+1}\dag+c_{i+1}) - h(c_i\dag c_i - \frac{1}{2}) \right. 
	\\ 	&\left. + J_z (c_i\dag c_i - \frac{1}{2})(c_{i+1}\dag c_{i+1} - 
	\frac{1}{2}) \right\rbrace	\\&\approx  \sum_{i=1}^{N} \left\lbrace 
	-\mu c_i\dag c_i +( t c_{i+1}\dag c_i + {\rm h.c.} )
	 - (\Delta c_{i+1}\dag c\dag_i + {\rm h.c.} ) \right\rbrace 
	 \\& \quad + {\rm const}, 
\end{split}
\end{equation}
where $\mu=h+J_z(1-2\braket{c\dag_{i}c_{i}})$, 
$t=1/4 - J_z\braket{c\dag_{i}c_{i+1}}$ and 
$\Delta=1/4-J_z\braket{c_{i}c_{i+1}}$ are determined self-consistently. 
These parameters are found to be real, and are shown in Fig.~\ref{fig7} 
as a function $h$, $J_z$ and $N$.

Using these self-consistent parameters, the Hamiltonian in 
Eq.~\eqref{eqHfermions} is then approximated by the following mean-field 
problem on an open chain:
\begin{equation} \begin{split} \label{eqHkit}
H_{\rm MF} &= -\sum_{i=1}^N \mu \left(c\dag_{i}c_{i}-\frac{1}{2}\right) 
\\&\quad + \sum_{i=1}^{N-1} \left[\left(t c\dag_{i+1}c_{i} + {\rm h.c.}\right) 
- \left( \Delta c\dag_{i+1}c\dag_{i} + {\rm h.c. }\right)   \right], 
	\end{split}
\end{equation}
up to an irrelevant additive constant~\footnote{In the periodic chain 
used to get the mean-field parameters, there is a level crossing when 
$\mu=2t$. To get good agreement with exact diagonalisation results and avoid a small 
discontinuity, we need to compute the expectation values in the state 
adiabatically connected to the ground state at $\mu<2t$. Thus for $\mu>2t$, 
the $\braket{.}$ are not computed in the ground state, but in the first 
excited state. Since all the level crossings arise for $\mu < 2t$, this has 
no influence on the following discussion.}.
\begin{figure}
{
	\includegraphics{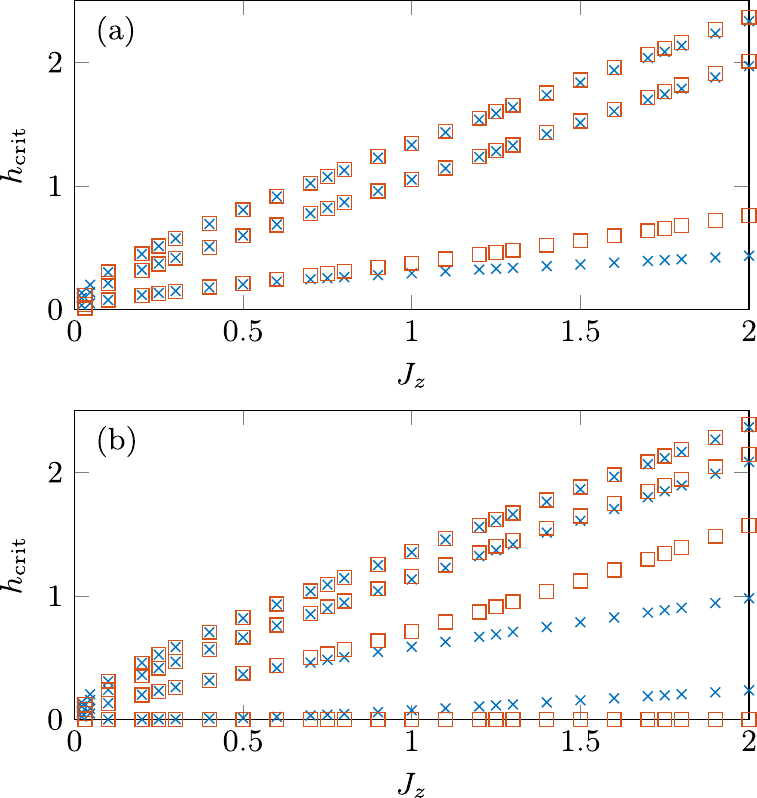}
}
\caption{Critical fields, $h_{\rm crit}$, where the degeneracy is exact, 
as a function of $J_z$ in the self-consistent mean-field approximation 
\eqref{eqHkit} (blue crosses) compared to the exact diagonalisation result 
(red squares) for (a) $N=6$ and (b) $N=7$.
}
\label{fig8}
\end{figure}
\begin{figure}[ht]
{
	\includegraphics{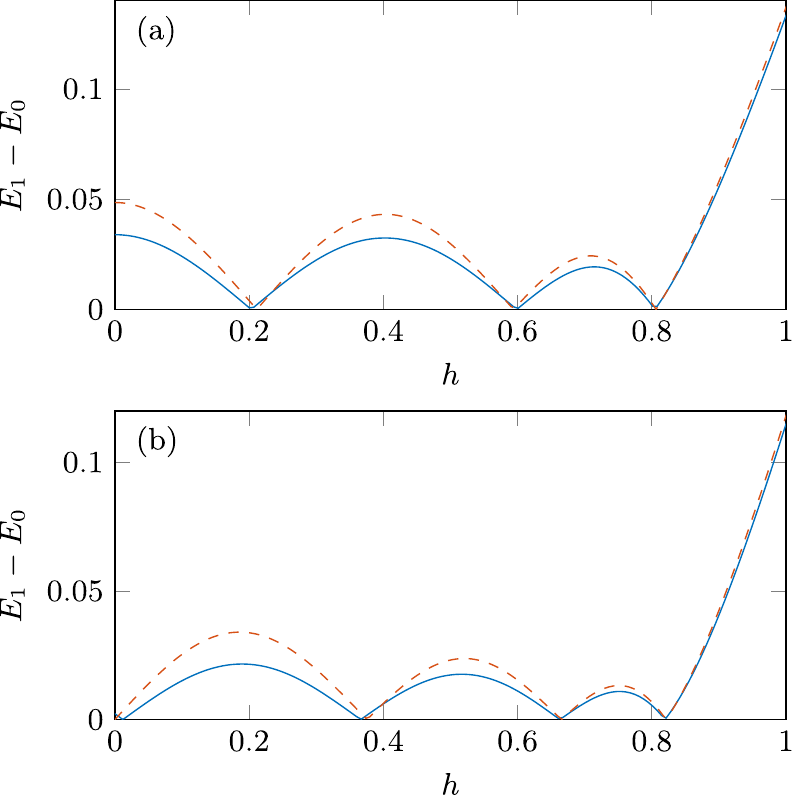}
}
\caption{Energy splitting $\epsilon=E_1-E_0$ as a function of $h$ in the 
self-consistent mean-field approximation \eqref{eqHkit} (blue solid line) 
compared to the exact diagonalisation result (red dashed line) for $J_z=0.5$
 and (a) $N=6$ and (b) $N=7$.
}
\label{fig9}
\end{figure}

Since the self-consistent parameters are almost independent of the system size 
(see Fig.~\ref{fig7}c), the boundaries are not very important and the bulk 
contribution is determinant. This partly justifies the approximation of 
playing with the boundary conditions to get the approximate model 
\eqref{eqHkit} with just three self-consistent parameters. This approximation 
is also justified by the great quantitative agreement with the exact 
diagonalisation results for the critical fields for $J_z \lesssim 0.8$ 
(see Fig.~\ref{fig8}), and to a lesser extent for the energy splitting 
$\epsilon=E_1-E_0$ between the two lowest energy states, see Fig.~\ref{fig9}. 

For $N$ odd, the degeneracy at $h=0$ is protected by symmetry for any $J_z$ 
in the Hamiltonian \eqref{eqHspin}. Indeed, under the transformation 
$S_i^z \to -S_i^z~\forall i$, the parity operator transforms as 
$P\to (-1)^N P$. Hence, for $N$ odd and $h=0$, the ground state has to be 
twofold degenerate. As can be seen in Fig.~\ref{fig8}b, the critical field 
$h=0$ at low $J_z$ evolves to a non-zero value for large $J_z$, thus showing 
that this symmetry is broken by the mean-field approximation \eqref{eqHkit}. 
The discrepancy is, however, small for $J_z \lesssim 0.8$ as can also be seen 
in Fig.~\ref{fig9}b. 

We observe from Fig.~\ref{fig7}a that as a function of magnetic field, the 
parameters $t$ and $\Delta$ are almost constant, whereas $\mu$ is almost 
proportional to $h$. Thus, we can understand the physics of the level 
oscillations by forgetting about the self-consistency and considering $\mu$, 
$t$ and $\Delta$ as free parameters, i.e. by studying the open Kitaev 
chain~\cite{kitaev}, where the level crossings happen as $\mu$ is tuned.
Compared to the TFI model for which $\Delta=t$, the main effect of $J_z>0$ 
is to make $0 < \Delta < t$, which, as we shall see in the next section, is 
the condition to see level oscillations.

Such a mapping between the two lowest lying energy states of the interacting 
Kitaev chain and of the non-interacting Kitaev chain can be made rigorous for a 
special value of $h>0$, provided the boundary terms in 
equation~\eqref{eqHfermions} are slighty modified~\cite{katsura}. But this 
particular exact case misses out on level-crossing oscillations.

\section{Level oscillations and Majorana fermions} \label{sec:Maj}
We define $2N$ Majorana operators $\gamma'_i,~\gamma_i''$ 
as: 
\begin{equation} \label{eqMaj}
	\begin{cases}
		\gamma'_{i} = c_{i}+ c\dag_{i}\\
		\gamma''_{i} = -\i (c_{i} -  c\dag_{i})
	\end{cases} 
\end{equation}
which satisfy ${\gamma'}_i\dag=\gamma'_i$, ${\gamma''}_i\dag=\gamma''_i$, 
$\lbrace \gamma'_i,\gamma''_j\rbrace=0$ and 
$\lbrace \gamma'_i,\gamma'_j\rbrace = 
\lbrace \gamma''_i,\gamma''_j\rbrace=2\delta_{ij}$. Since the $\mu,~t,~\Delta$ 
are real, the $H_{MF}$ of Eq.~\eqref{eqHkit} reads
\begin{equation}
	\begin{split}
H_{\rm MF}=& \frac{\i}{2}\sum_{i=1}^{N-1} 
\left[ -(t+\Delta)\gamma''_{i}\gamma'_{i+1} 
+ (t-\Delta )\gamma'_{i}\gamma''_{i+1} \right] \\
& -\frac{\i \mu}{2} \sum_{i=1}^N \gamma'_{i}\gamma''_{i}= 
\frac{\i}{2} \sum_{i,j=1}^{N} \gamma'_iM_{ij}\gamma''_j.
	\end{split}
\end{equation}
From the singular value decomposition of $M$, we write $M=U\Sigma V^T$, 
where $U$ and $V$ are orthogonal matrices and 
$\Sigma={\rm diag}(\epsilon_1,\ldots , \epsilon_N)$ with real $\epsilon_i$ 
and $|\epsilon_i|\leq |\epsilon_{i+1}|~\forall i$. Thus, the Hamiltonian reads
\begin{equation}
\begin{split}
H_{\rm MF} &= \frac{\i}{2} \sum_{i,j,k=1}^{N} 
\gamma'_i U_{ik}\epsilon_k V^T_{kj}\gamma''_j
=\frac{\i}{2} \sum_{k=1}^{N} \epsilon_k \tilde \gamma'_k \tilde \gamma''_k \\
&=\sum_k \epsilon_k (\eta\dag_{k}\eta_{k}-\frac{1}{2})
\end{split}
\end{equation}
where
\begin{equation}
	\tilde \gamma'_k = \sum_{i=1}^N \gamma'_i U_{ik}, \qquad \tilde 
	\gamma''_k = \sum_{i=1}^N \gamma''_i V_{ik}
\end{equation}
are the rotated Majorana operators, and the 
$\eta_k = \frac{1}{2} (\tilde \gamma'_{k}+\i \tilde \gamma''_{k}) $ are 
fermionic annihilation operators corresponding to the Bogoliubov 
quasiparticles.

 As derived in Appendix, in general the Majorana operators, $\tilde \gamma'_k$ 
 and $\tilde\gamma''_k$, are of the form
 \begin{equation} \label{eqtildegammageneral}
 \begin{split}
		\tilde \gamma'_k = \sum_j (a_+ x_+^j + b_+ x_+^{N+1-j} 
		+ a_- x_-^j + b_- x_-^{N+1-j} ) \gamma'_{j}\\
		\tilde \gamma''_k = \sum_j ( a_+ x_+^{N+1-j} + b_+ x_+^j 
		+a_- x_-^{N+1-j} + b_- x_-^j) \gamma''_j{}
 \end{split}
 \end{equation}
 where the $x_\pm$, $a_\pm$ and $b_\pm$ are functions of the energy 
 $\epsilon_k$ which is quantised in order to satisfy the boundary 
 conditions. On can easily solve numerically the nonlinear equation for 
 the $\epsilon_k$. Here, we will instead focus on a simple analytical 
 approximation for $\tilde \gamma'_1,~\tilde\gamma''_1$ and $\epsilon_1$ 
 which works well to discuss the level crossings,
 and is equivalent to the Ansatz given in \cite{kitaev}.

From Eqs. \eqref{eqratio} and \eqref{eqEps}, we see that for $\epsilon=0$, 
we have either $a_\pm=0$ or $b_\pm=0$. Without loss of generality, we can 
choose $b_\pm(\epsilon=0) = 0$. Since we expect $\epsilon_1 \ll 1$, we 
approximate 
 \begin{equation} \label{eqapp1}
 	b_\pm (\epsilon_1) \approx b_\pm (0)=0
 \end{equation}
 and 
 \begin{equation}\label{eqapp2}
 	x_\pm (\epsilon_1) \approx x_\pm(0) = 
 	\frac{\mu \pm \sqrt{\mu^2-4t^2+4\Delta^2}}{2(t+\Delta)},
 \end{equation}
which yields
  \begin{equation}  \label{eqapp3}
 \begin{split}
\tilde \gamma'_1 &\approx \sum_j (a_+ x_+^j  + a_- x_-^j ) \gamma'_{j}\\
\tilde \gamma''_1 &\approx\sum_j ( a_+ x_+^{N+1-j} +a_- x_-^{N+1-j} )\gamma''_j
 \end{split}
 \end{equation}
 with $\sum_j (a_+ x_+^j  + a_- x_-^j )^2=1$.
 The boundary conditions \eqref{eqbc} now read 
 \begin{subequations} \label{eqbc12}
\begin{eqnarray}
 a_+ + a_- &=& 0 \label{eqbc1} \\
	 a_+ x_+^{N+1} + a_- x_-^{N+1} &=& 0 \label{eqbc2}
\end{eqnarray}
\end{subequations}
and in general cannot be both satisfied unless $\epsilon_1=0$ exactly. 

If $|x_\pm|<1$, $\tilde\gamma'_1$ is localised on the left side of the chain 
with its amplitude $\sim e^{-j/\xi}$ as $j\gg 1$ with 
$\xi=-1/\ln (\max(|x_+|,|x_-|))$. Furthermore, $\tilde\gamma''_1$ is related 
to $\tilde\gamma'_1$ by the reflection symmetry $j\to N+1-j$.
Thus, in the thermodynamic limit, the boundary condition \eqref{eqbc2} is 
irrelevant and $\epsilon_1 \to 0$ as $N\to \infty$. 
Similarly, if $|x_\pm| >1$ the boundary condition \eqref{eqbc1} becomes 
irrelevant in the thermodynamic limit.
However, if $|x_+|>1$ and $|x_-|<1$,  or $|x_+|<1$ and $|x_-|>1$, then 
$\tilde\gamma'_1$, $\tilde\gamma''_1$ have significant weight on both 
sides of the chain and both boundary conditions \eqref{eqbc1} and 
\eqref{eqbc2} remain important in the thermodynamic limit. Hence, the 
approximation $\epsilon_1 \approx 0$ is bad, indicating a gapped system.
As discussed in \cite{kitaev}, for $|\mu|<2|t|$ we have either $|x_\pm|<1$ 
or $|x_\pm|>1$ which yields $\epsilon_1=0$ in the thermodynamic limit. This 
is the topological phase with a twofold degenerate ground state. For a finite 
system, however, the boundary conditions \eqref{eqbc1} and \eqref{eqbc2} are 
in general not exactly satisfied and the system is only quasi-degenerate with 
a gap $\epsilon \sim e^{-N/\xi}$.
For $|\mu|>2|t|$, either $|x_+|>1$ and $|x_-|<1$, or $|x_+|<1$ and 
$|x_-|>1$, and the system is gapped. 

\begin{figure}[t]
{
	\includegraphics{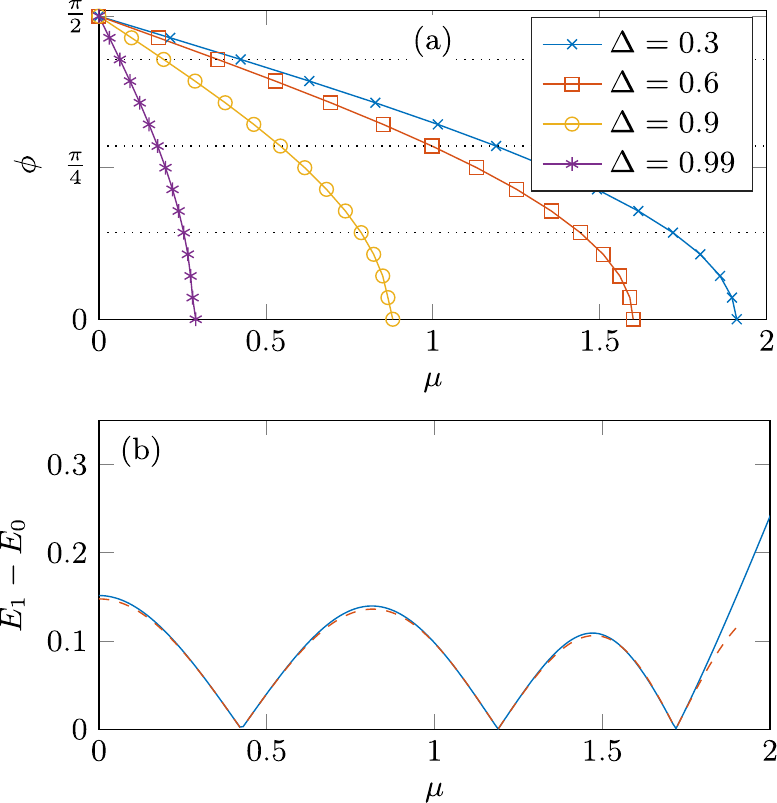}
}
\caption{(a) Phase $\phi(\mu)$ of $x_+=re^{\i\phi}$ within the approximation 
\eqref{eqapp2} for several $\Delta$ with $t=1$ and $N=6$. The horizontal black 
dotted lines indicate the values $\phi=\frac{\pi m}{N+1}$. (b) Splitting 
$E_1-E_0=|\epsilon_1|$ in the Kitaev chain calculated exactly solving 
numerically the full self-consistent equations described in Appendix 
(blue solid line) and with the analytical approximate result 
in Eq.~\eqref{eqapprox3} (red dashed line) for $N=6$, $t=1$ and $\Delta=0.3$.
}
\label{fig10}
\end{figure}

In the topological phase, $|\mu|<2|t|$, there are parameters for which the 
boundary conditions \eqref{eqbc12} can be exactly satisfied even for 
$N<\infty$ and thus $\epsilon_1=0$ exactly. In such a case, there is an 
exact zero mode even for a finite chain. This was previously discussed in 
Ref.~\cite{kao}, as well as in \cite{hedge} where a more general method that
applies to disordered systems is described. If $x_\pm \in \mathbb{R}$, it 
is never possible to satisfy the boundary conditions \eqref{eqbc12} and 
therefore the quasi-gap is always finite, $\epsilon_1 \neq 0$. However, if 
$x_+=re^{\i\phi} \not \in \mathbb{R}$, Eq.~\eqref{eqapp2} yields $x_-=x_+^*$ 
and $(x_+^{N+1} - x_-^{N+1}) \propto r^{N+1}\sin[(N+1)\phi]$. 
Thus it may happen for specific parameters that $\epsilon_1=0$ exactly. This 
degeneracy indicates a level crossing. The phase $\phi$, defined for 
$|\mu|<\mu_c=2\sqrt{t^2-\Delta^2}$, is given by
\begin{equation}
	\tan{\phi}  = \sqrt{(\mu_c/\mu)^2-1}.
\end{equation}
It thus goes continuously from $\phi(\mu=0^+)=\pi/2$ to 
$\phi(\mu\to\mu_c) \to 0$. Hence, there are critical chemical potentials, 
$0 \leq \mu_{\lceil N/2\rceil} <  \ldots < \mu_m <\ldots<\mu_1<\mu_c$, 
such that $\phi(\mu=\mu_m)=\frac{\pi m}{N+1}$ (see Fig.~\ref{fig10}a). 
For these critical $\mu_m$, the system is exactly degenerate, i.e. 
$\epsilon_1=0$. In the TFI limit, we have $\Delta=t$ and $\mu_c=0$, 
thus there are no level crossings.

For $|\mu|<2|t|$, writing $x_+=re^{\i \phi}$ with $r>0$, we have
\begin{equation} \label{eqapprox3}
	\begin{split}
		\epsilon_1 &=\Sigma_{11}=(U^T M V)_{11}  \\
		&\approx 4(t+\Delta) a_+^2 r^{N+2} \sin(\phi) \sin[(N+1)\phi], 
	\end{split}
\end{equation}
where we used the approximations \eqref{eqapp2}, \eqref{eqapp3} and the 
boundary condition \eqref{eqbc1} [respectively \eqref{eqbc2}] when 
$t\Delta>0$ (respectively $t\Delta<0$), since in this case $|x_\pm|<1$ 
(respectively $|x_\pm|>1$). Note that $\phi(-\mu) = \phi(\mu)-\pi$, 
and thus $\epsilon_1$ is an odd function of $\mu$ for odd $N$ and an 
even function of $\mu$ for even $N$. Since $\epsilon_1$ changes sign 
whenever $\sin((N+1)\phi)=0$, the degeneracy points indicate level crossings. 
This approximate description works extremely well, as shown in 
Fig.~\ref{fig10}b for $\Delta=0.3t$. Because $\phi$ takes all the values 
in $]0,\pi/2]$ for $0 < \mu <\mu_c$, and in $]-\pi,-\pi/2]$ for 
$-\mu_c < \mu < 0$, there are either exactly $N$ level crossings as a function 
of $\mu$ if $0<\mu_c\in \mathbb{R}$, i.e. if $|\Delta|<|t|$, and no zero 
level crossing otherwise. At the points of exact degeneracy, 
$b_\pm(\epsilon=0)=0$, the zero-mode Majorana fermions are localised on 
opposite sides of the chain. When the degeneracy is not exact, however, 
$b_\pm (\epsilon\neq 0) \neq 0$ and the zero-mode Majorana fermions mix 
together to form Majoranas localised mostly on one side but also a little 
bit on the opposite side.

In the XY model in an out-of-plane magnetic field, which is equivalent to 
the non-interacting Kitaev chain~\cite{lieb}, these level crossings lead 
to an oscillatory behaviour of the spin correlation functions~\cite{barouch}. 
In the context of p-wave 
superconductors, the level oscillations described above also 
arise in more realistic models and are considered one of the hallmarks of 
the presence of topological Majorana fermions~\cite{sarma,loss}. Although 
it is still debated whether Majorana fermions have already been observed, 
strong experimental evidence for the level oscillations was reported in 
\cite{markus}.

\begin{figure}
{
	\includegraphics{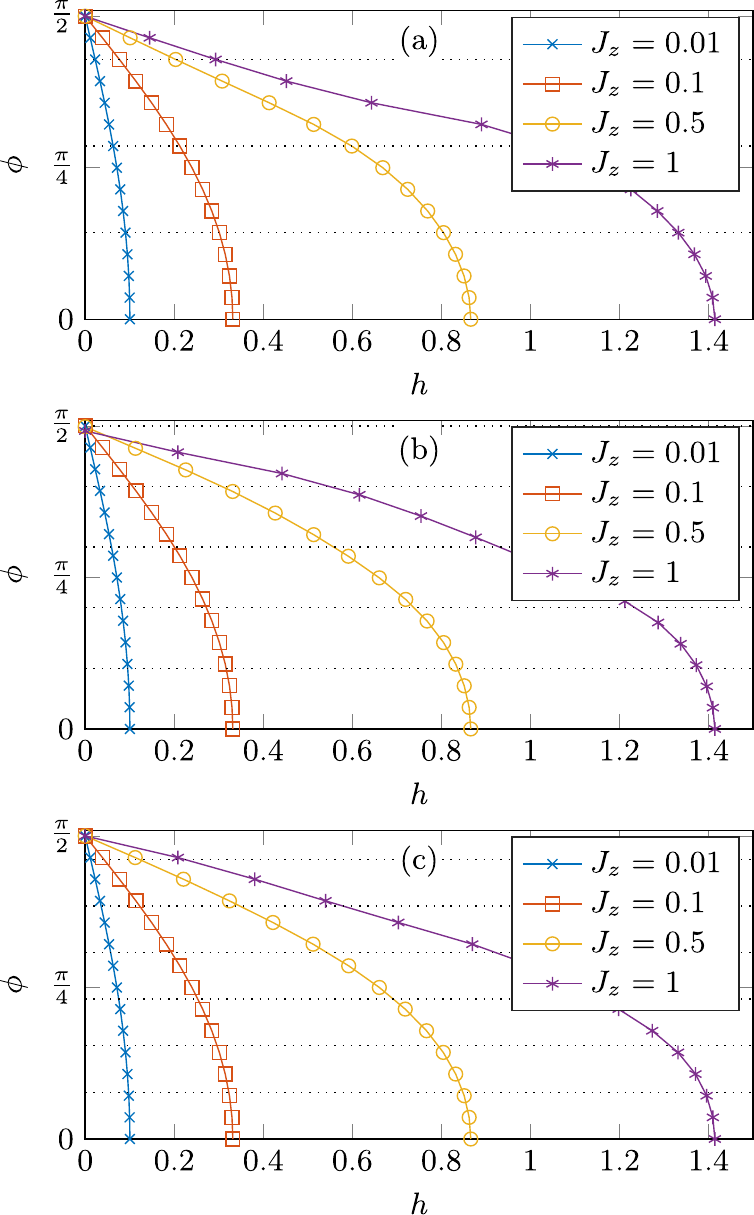}
}
\caption{Phase $\phi(h)$ of $x_+(\epsilon=0)$ based on the self-consistent 
parameters $\mu,~t,~\Delta$ of the mean-field decoupling for several $J_z$ 
and (a) $N=6$, (b) $N=9$, (c) $N=12$. The horizontal black dotted lines
indicate the values $\phi=\frac{\pi m}{N+1}$.
}
\label{fig11}
\end{figure}

Coming back to the mean-field Hamiltonian of Eq.~\eqref{eqMF}, we can get 
the phase $\phi$ within the approximation \eqref{eqapp2}, i.e. the phase of 
$x_+(\epsilon=0)$, as a function of the physical parameters $h,~J_z$ since 
we know how the self-consistent parameters $\mu,~t,~\Delta$ depend on them.
We plot in Fig.~\ref{fig11} the phase $\phi$ as a function of $h$ for 
several $J_z$ which yields a good qualitative understanding of the sudden 
appearance of $N$ level crossings as soon as $J_z > 0$. As previously 
discussed, the self-consistent parameters are almost independent of $N$ and 
therefore the curves $\phi(h)$ are almost independent of $N$ as well. The 
main effect of $N$ is to change the condition 
$\phi(\mu=\mu_m)=\frac{\pi m}{N+1}$ for the boundary condition in 
Eq.~\eqref{eqbc2} to be satisfied and thus for the system to be exactly 
degenerate.

\section{Summary}\label{sec:exp}

The main result of this paper is that the level crossings between the two 
lowest energy eigenstates of the XY chain in an in-plane magnetic field 
are more generally a fundamental feature of the transverse field Ising chain 
with an antiferromagnetic longitudinal coupling  
howsoever small. These points of level crossings (twofold degeneracy) 
correspond to having Majorana edge modes in a Kitaev chain onto which the 
problem can be approximately mapped. The level crossings of the XY chains 
have been observed experimentally in \cite{toskovic} by scanning tunneling 
microscopy on Cobalt atoms evaporated onto a Cu${}_2$N/Cu(100) substrate. 
By varying the adsorbed atoms and the substrate, it should be possible to 
vary the easy-plane and easy-axis anisotropies, and thus to explore the 
exact degeneracy points for various values of the longitudinal coupling.
The possibility to probe the two-fold degeneracy of this family of spin 
chains is important in view of their potential use for universal quantum
computation~\cite{loss2}.
Besides, one could also realise the spinless fermionic Hamiltonian 
\eqref{eqHfermions} in an array of Josephson junctions as described in 
\cite{hassler}. The advantage of this realisation is that it allows a 
great flexibility to tune all the parameters of the model. We hope that 
the results of the present paper will stimulate experimental investigations 
along these lines.

\begin{acknowledgments}
We acknowledge Somenath Jalal for useful discussions and the Swiss National 
Science Foundation for financial support. B.K. acknowledges the financial 
support under UPE-II and DST-PURSE programs of JNU.
\end{acknowledgments}

\appendix*

\section{Majorana solutions of the Kitaev chain}\label{secMajWF}

To solve the Kitaev chain \eqref{eqHkit}, we need to find the singular 
value decomposition of
\begin{equation}
	M=\begin{pmatrix}
		-\mu & \tau_- & 0  & \cdots\\
		\tau_+ & -\mu & \tau_- & 0 &\cdots \\
		0 &\tau_+ & -\mu & \tau_- & 0 & \cdots \\
		 & & \ddots  & \ddots & \ddots & \\
		&\cdots & 0 & \tau_+ & -\mu & \tau_- \\
		& & \cdots & 0 & \tau_+ & -\mu
	\end{pmatrix}
\end{equation}
with $\tau_\pm = t\pm\Delta$, i.e. find orthogonal matrices $U$, $V$ 
and a real diagonal matrix $\Sigma$ such that $M=U\Sigma V^T$. Writing 
$\vec u_k$ and $\vec v_k$ the $k^{\rm th}$ columns of $U$ and $V$ 
respectively, they satisfy
\begin{equation} \label{eqsvd}
	\begin{cases}
		M\vec v_k &= \epsilon_k \vec u_k \\
		\vec u_k^T M &= \epsilon_k \vec v_k^T.
	\end{cases}
\end{equation}

Let's find two unit-norm column vectors $\vec u,~\vec v$ and $\epsilon$ 
such that $M\vec v=\epsilon \vec u$ and $\vec u^TM=\epsilon \vec v^T$. 
First we forget about the normalisation and boundary conditions and focus 
on the secular equation. 
Setting the components of $\vec u,~\vec v$ as $u_j=a x^j$ and $v_j = b x^j$, 
we have
\begin{equation}
	\begin{split}
	M \vec v &= \frac{b}{a}\frac{\tau_+ -\mu x +\tau_- x^2}{x}  
	\vec u+ {\rm b.t.} \\
	\vec u^T M &=\frac{a}{b}\frac{\tau_- -\mu x +\tau_+ x^2}{x} 
	\vec v^T + {\rm b.t.}
	\end{split}
\end{equation}
where b.t. stands for boundary terms. Hence, $u$ and $v$ satisfy the secular 
equation provided 
\begin{equation} \label{eqratio}
	\frac{b}{a}=\sqrt{\frac{\tau_- -\mu x +\tau_+ x^2}
	{\tau_+ -\mu x +\tau_- x^2}}
\end{equation}
and
\begin{equation} \label{eqEps}
	\epsilon = \frac{1}{x}\sqrt{(\tau_- -\mu x +\tau_+ x^2)
	(\tau_+ -\mu x +\tau_- x^2)}.
\end{equation}
Because of the reflection symmetry $j \to N+1-j$, if $x$ is a solution of 
equation \eqref{eqEps} for some $\epsilon$, then $1/x$ is also a solution. 
Assuming $\epsilon$ known, the solutions are $x_\pm$, $1/x_\pm$ and satisfy
\begin{equation}
	\begin{split}
		0 &= \epsilon^2 x^2 - (\tau_- -\mu x +\tau_+ x^2)
		(\tau_+ -\mu x +\tau_- x^2) \\
		&\propto (x-x_+)(x-1/x_+)(x-x_-)(x-1/x_-)
	\end{split}
\end{equation}
which by identification yields, writing $\rho_\pm = x_\pm + 1/x_\pm$,
\begin{equation} \label{eqxpm}
	\begin{split} 
		x_\pm &= \frac{1}{2}\left(\rho_\pm +\sqrt{\rho_\pm^2 - 4}\right),\\
		\rho_\pm &= \frac{\mu t \pm \sqrt{(t^2-\Delta^2)\epsilon^2 + 
		\Delta^2(\mu^2-4t^2+4\Delta^2)}}{t^2-\Delta^2}.
	\end{split}
\end{equation}
Taking into account the reflection symmetry, the general form of the 
components of $\vec u,~\vec v$ is thus
 \begin{equation} \label{equv}
\begin{split}
	u_j = a_+ x_+^j + b_+ x_+^{N+1-j} + a_- x_-^j + b_- x_-^{N+1-j} \\
		v_j =  a_+ x_+^{N+1-j} + b_+ x_+^j +a_- x_-^{N+1-j} + b_- x_-^j
\end{split}
 \end{equation}
with the ratios $b_+/a_+$ and $b_-/a_-$ given by equation \eqref{eqratio} 
with $x=x_+$ and $x=x_-$ respectively.

Furthermore, we have the boundary conditions
\begin{equation} \label{eqbc}
	\begin{split}
		a_+ + b_+x_+^{N+1} + a_- + b_-x_-^{N+1}&= 0 \\
	a_+ x_+^{N+1} + b_+ + a_-x_-^{N+1} + b_-&= 0
	\end{split}
\end{equation}
which set the ratio $a_-/a_+$ and give the quantisation condition on the 
energies $\epsilon_k$.
The last degree of freedom, say $a_+$, is then set by normalising $\vec u$ 
(from equation \eqref{equv}, $\| \vec u\| = \| \vec v \|$).

Note that for the special cases $t= \Delta$ and $\mu=0$, we have 
$\tilde \gamma'_1 = \gamma'_1$ and $\tilde \gamma''_1=\gamma''_N$ 
with $\epsilon_1=0$. We have a similar result for $t=-\Delta$ and $\mu=0$. 
For these two cases, the general formalism described above does not apply 
since it yields $x_\pm=0,\pm\infty$.

\end{document}